\documentclass[a4paper,12pt]{article}
\usepackage{psfig}
\usepackage{latexsym}

\def\gsim {\mbox{\hbox{ \lower-.6ex\hbox{$>$}
\kern-1.12em \lower.5ex\hbox{$\sim$}\kern+.35em}}}
\def\lsim {\mbox{\hbox{ \lower-.6ex\hbox{$<$}
\kern-1.12em \lower.5ex\hbox{$\sim$}\kern+.35em}}}

\def\Ob{{\bf O}}
\def\Rb{{\bf R}}

\def\Kb{{\bf K}}

\def\kb{{\bf k}}
\def\vb{{\bf v}}

\def\rb{{\bf r}}
\def\gamdot{{\dot \gamma}}
\begin{document}

\title{\textbf{Fluid Vesicles in Shear Flow}}
\author{Martin Kraus, Wolfgang Wintz, Udo Seifert and Reinhard Lipowsky
\\ \\  
Max-Planck-Institut f\"ur Kolloid- und Grenzfl\"achenforschung\\
Kantstr. 55, 14513 Teltow-Seehof, Germany\\
}

\date{\today}

\maketitle
\begin{abstract} 
  The shape dynamics of fluid vesicles is governed by the coupling of
  the flow within the two-dimensional membrane to the hydrodynamics of
  the surrounding bulk fluid. We present a numerical scheme which is
  capable of solving this flow problem for arbitrarily shaped vesicles
  using the Oseen tensor formalism.  For the particular problem of
  simple shear flow, stationary shapes are found for a large range of
  parameters.  The dependence of the orientation of the vesicle and
  the membrane velocity on shear rate and vesicle volume can be
  understood from a simplified model.
\end{abstract}


\vskip 1cm


Vesicles are closed lipid membranes suspended in aqueous solution.  If
the composition of this solution inside and outside the vesicle is
identical, it affects the equilibrium properties of the vesicle only
by osmotically fixing the enclosed volume. Minimizing the bending
energy of the membrane under the constraints of fixed enclosed volume
and membrane area then yields the equilibrium vesicle shape at
rest\cite{seif95b}.  The dynamics of this shape, however, is governed
by the coupling of the flow within the two-dimensional incompressible
fluid membrane to the hydrodynamics of the bulk fluid.  Any theory of
vesicle dynamics is complicated by the fact, that the boundary
conditions for the three-dimensional Navier-Stokes equations have to
be evaluated at the vesicle surface, which is moving with the fluid
and whose shape is not known a priori.

Considerable practical interest in this problem arises from a more
complex system, the red blood cell.  The deformation of erythrocytes
in hydrodynamic flow fields is used as a measure of changes in the
elastic properties of pathologically modified cells
\cite{fisc78,gron80}. The understanding of the dynamics of a single
cell is also a prerequisite for the understanding of the rheology of
blood.  Since the erythrocyte has a complicated structure consisting
of a fluid lipid/protein bilayer and the spectrin network, theoretical
analysis has employed expansions for small flow rate \cite{pete92b} or
has focused on simpler model systems such as shells with stretching
elasticity and liquid drops with homogeneous surface tension
\cite{kell82,bart91,uijt93,pozr95}.  The dynamics of fluid membranes
with fixed area and bending rigidity has only been studied in the
quasiplanar \cite{broc75a}, quasispherical \cite{miln87} and
cylindrical \cite{nels95} geometry without external flow fields.
Further analytical work includes the covariant treatment of the
membrane incompressibility \cite{folt94} and the dynamic
renormalization of material parameters \cite{cai95} in equilibrium.
The problem for a closed vesicle with arbitrary shape and external
flow is investigated in this Letter for the first time.


\begin{figure}[b]
  \begin{minipage}{6.1cm}
  \begin{center}
    \leavevmode
    \psfig{file=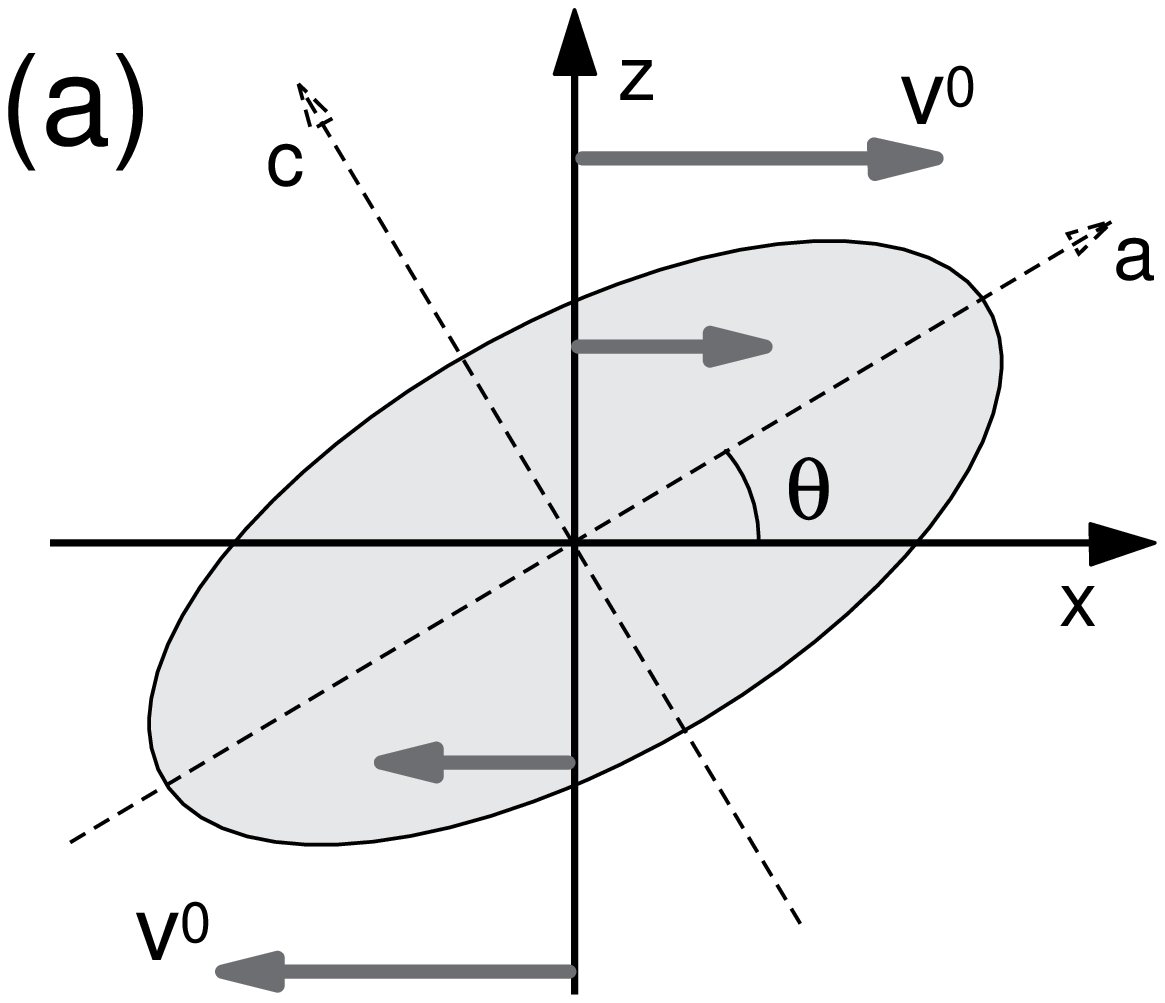,width=6cm}
  \end{center}
  \end{minipage}
  \begin{minipage}{7.6cm}
  \begin{center}
    \leavevmode
    \psfig{file=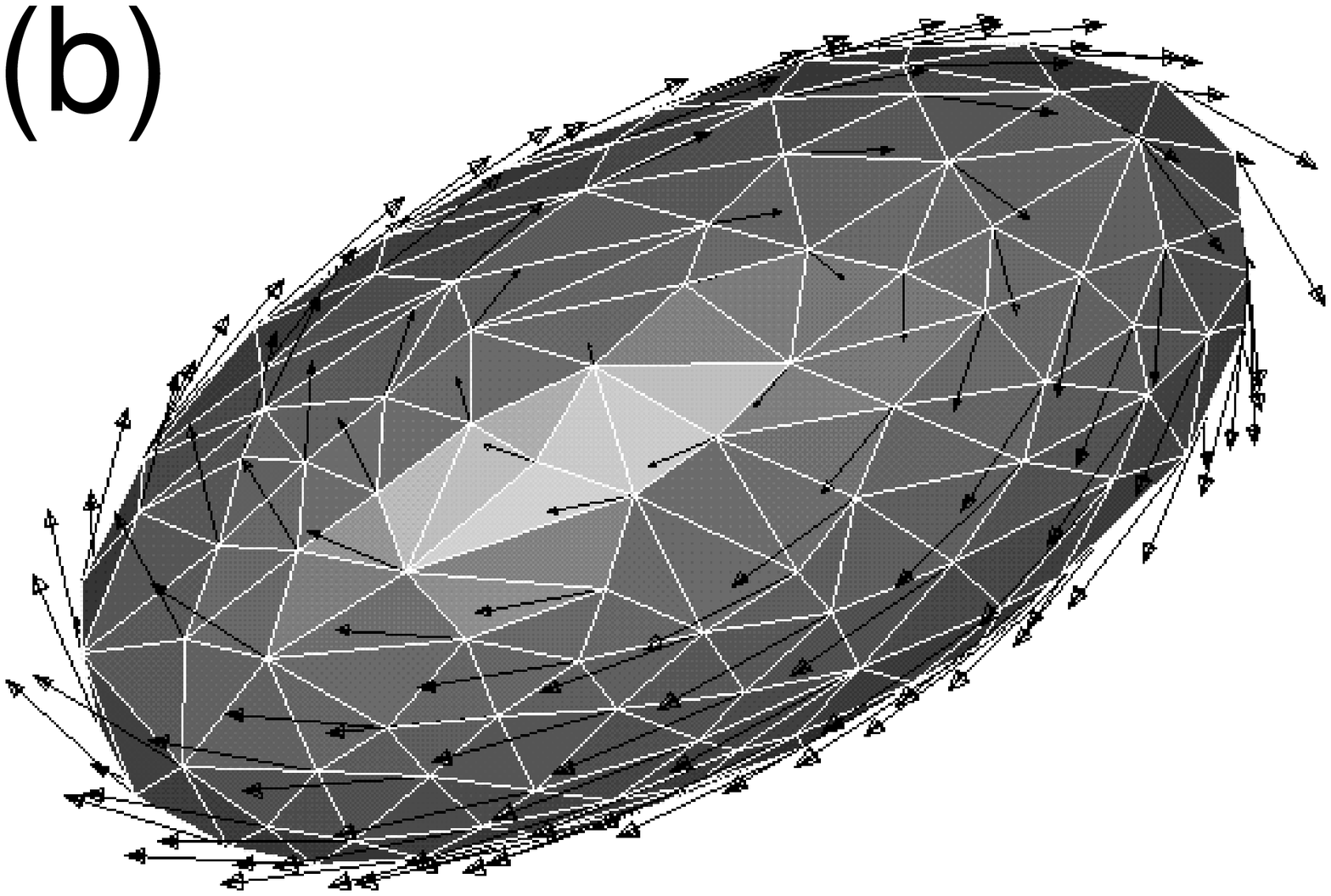,width=6.8cm}
  \end{center} 
  \end{minipage}
  \caption{Vesicle in shear flow $v^0_x = \gamdot z$. (a)
    Schematic drawing of the undisturbed flow, inclination angle
    $\theta$, and coordinate axes $x$, $z$, $a$ and $c$.  The
    coordinate axis pointing into the paper plane is $y$.  (b)
    Stationary state obtained by numerical integration for $v=0.9$,
    $\gamdot = 1$ s$^{-1}$, $\kappa = 10^{-19}$ J, $\eta = 10^{-3}$ Js
    / m$^{-3}$, and $R_0 = 10 \mu$m.  The velocities at the vertices
    are shown as arrows with length proportional to velocity.}
  \label{fig:shearscheme}
\end{figure}

\begin{figure}[p]
  \begin{center}
    \leavevmode
    \psfig{file=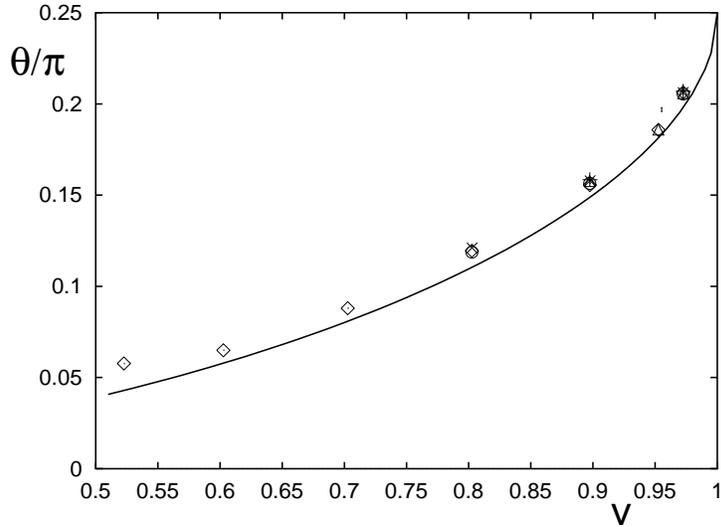,width=9.5cm}
  \end{center}
  \caption{
    Inclination angle $\theta$ between the $x$-axis and the longest
    axis of inertia as a function of reduced volume $v$ for different
    reduced shear rates $\chi = $1 ($\bigtriangleup$), 5 ($\bigcirc$),
    10 ($\Diamond$), 50 ($\times$), 100 ($+$).  Numerical errors are
    smaller than the symbol sizes. The continuous line follows from
    Eq.~(\ref{shear_ellips}).}
  \label{fig:angle}
\end{figure}

\begin{figure}[p]
  \begin{center}
    \leavevmode
    \psfig{file=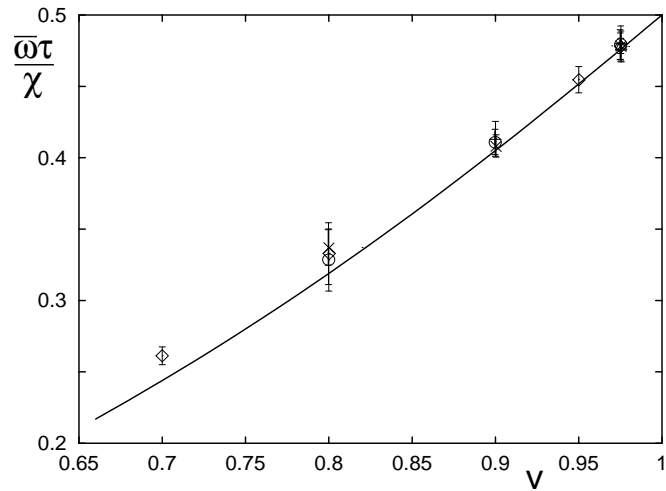,width=9.5cm}
  \end{center}
  \caption{
    Scaled revolution frequency $\overline{\omega} \tau / \chi$
    averaged over all vertices as a function of reduced volume $v$ for
    different reduced shear rates $\chi$ (Symbols are as in
    Fig.~\ref{fig:angle}).  The continuous line is computed using the
    solution of (\ref{angle_dgl_beta}). For volumes $v < 0.7$, a full
    revolution in the stationary state could not be computed due to
    constraints on computation time.}
  \label{fig:tt}
\end{figure}

For specifity, we investigate a linear shear flow $v^0_x(\rb) =
\gamdot z$ with shear rate $\gamdot$, as sketched in
Fig.~\ref{fig:shearscheme} (a), which can easily be realised
experimentally \cite{fisc78}.  The stationary state of a vesicle in
such a flow resulting from our computation is shown in
Fig.~\ref{fig:shearscheme} (b). It is characterized by both a finite
inclination angle $\theta$ between the longest axis of inertia of the
vesicle and the flow direction, and a 'tank-treading' tangential
motion of the membrane around the vesicle with revolution frequency
$\omega$.  Both $\theta$ and $\omega$ are found to depend strongly on
the reduced volume $v = V / (4 \pi R_0^3 / 3)$ of the vesicle, where
$V$ is the enclosed volume and the surface area $A$ determines the
length-scale $R_0 = \sqrt{A / 4 \pi}$. The bending rigidity of the
membrane $\kappa$ sets the energy scale \cite{fnt_c0}. Together with
the bulk viscosity $\eta$, these quantities define an intrinsic
timescale $\tau \equiv \eta R_0^3 / \kappa$.  Thus, the dimensionless
shear rate $\chi \equiv \gamdot \tau = \gamdot \eta R_0^3 / \kappa$ is
the second parameter of the problem.  Figures
\ref{fig:angle} and \ref{fig:tt} show the inclination angle $\theta$
and the mean revolution frequency $\overline{\omega}$
\cite{fnt_numtt}, respectively, as a function of the reduced volume
$v$ for different values of $\chi$.

We find numerically that for any non-zero shear rate small
perturbations of the equilibrium shape do not allow stationary flow
with the appropriate boundary conditions \cite{krau96}. Thus, shear
appears to be a singular perturbation that drives the vesicle towards
a stationary non-equilibrium shape that can be distinctly different
from the corresponding equilibrium shape.  Further increasing the
shear rate $\chi$ only marginally affects shape and orientation angle
$\theta$.
This behavior is
remarkably different from the behavior of both erythrocytes, which go
through a transition from tumbling to tanktreading behavior with
increasing $\gamdot$ \cite{fisc78}, and liquid drops, which deform to
more and more elongated shapes as a function of $\gamdot$ until they
break up \cite{rall84}.

The stationary state was found to be independent of initial
conditions.  In particular, for $v \lsim 0.75$, oblate discocytes are
locally stable in the absence of an external flow field \cite{krau95z}
but still transform to the same elongated shape as prolate vesicles
when suspended in shear flow. We have tested this behavior for all $v
\geq 0.52$, i.e.~the entire range where the equilibrium discocytes do
not self-intersect.


Hydrodynamic calculations on typical length and time scales of
membranes and vesicles are simplified by employing the Stokes
approximation which is valid in the limit of small Reynolds number
${\sf Re}$ \cite{vdve89}. For a vesicle suspended in a shear flow with
typical velocity $u = \gamdot R_0$, this number can be estimated by
${\sf Re} = \rho u R_0 / \eta = \rho \gamdot R_0^2 / \eta$.  Inserting
the vesicle size $R_0 \simeq 10 \mu$m and the viscosity $\eta =
10^{-3} {\rm Js} / {\rm m}^3$ and density $\rho = 10^3 {\rm kg} / {\rm
  m}^3$ of water, the limit of small ${\sf Re}$ corresponds to
$\dot\gamma \ll 10^4 {\rm s}^{-1}$.  Thus, the hydrodynamic equations
for the velocity field $\vb(\rb)$ of an incompressible Newtonian bulk
fluid with external force density $\kb(\rb)$ reduce to the
instantaneous force balance
\begin{equation}
  \eta \nabla^2 \vb=\nabla p(\rb) + \kb (\rb).
\label{stokeshydro}
\end{equation}
The pressure field $p(\rb)$ can be eliminated by using the
incompressibility condition \mbox{$\nabla \cdot \vb(\rb) =0$}.  Due to
the linearity of (\ref{stokeshydro}), $\vb(\rb)$ can be written as a
superposition of a background flow $\vb^0(\rb)$ and additional flows
$\vb^i(\rb)$ arising from the interactions of the fluid with the
membrane.


We describe the membrane surface by a triangulation with vertex
positions $\Rb^\alpha$ moving with the adjacent bulk fluid due to
non-slip boundary conditions and impermeability of the membrane for
bulk flow.  Forces on the membrane are computed on these vertices from
the energy
\begin{equation}
  \label{helf_ham_discr}
  {\cal G}(\{\Rb^\alpha\}) = \sum_\alpha  2\kappa (H^2)^\alpha
  + \sum_\alpha \sigma^\alpha A^\alpha
\end{equation} 
of the discretized membrane. The energy has two contributions. First,
the discretized squared mean curvature $(H^2)^\alpha$
\cite{fnt_curvdiscr} describes the bending energy
\cite{canh70,helf73,fnt_c0} with bending rigidity $\kappa$. Second, a
locally varying isotropic tension $\sigma^\alpha$  which is
conjugate to the area $A^\alpha$ of the neighboring triangles around
each vertex is needed to ensure local incompressibility of the
membrane.

The force $\Kb^\beta$ at vertex $\Rb^\beta$ then reads
\begin{eqnarray}
\label{force}
\Kb^\beta 
& \equiv & - \frac{\partial {\cal G}(\{\Rb^\alpha\})}
{\partial \Rb^\beta} \nonumber \\ 
& = &  - 2 \kappa\sum_\alpha \frac{\partial
  (H^2)^\alpha}{\partial \Rb^\beta} 
  - \sum_\alpha\sigma^\alpha \frac{\partial
  A^\alpha}{\partial \Rb^\beta},
\end{eqnarray}
and the force density is given by $\kb(\rb) = \sum_\beta \delta(\rb -
\Rb^\beta) \Kb^\beta$.

The disturbance flow due to the presence of the vesicle can be
computed using the Oseen-tensor $\Ob$ describing the flow field
induced by a discrete point force $\Kb^\beta$ in an infinite
fluid medium. The total velocity at
the vertex position $\Rb^\alpha$ then reads
\begin{equation}
  \vb(\Rb^\alpha) = \vb^0(\Rb^\alpha) + \sum_\beta 
    \Ob(\Rb^\alpha, \Rb^\beta) \Kb^\beta.
  \label{veloc}
\end{equation}
The matrix elements of the Oseen-tensor are
\cite{pozr92,doi86,fnt_oseendiv}
\begin{equation}
  \Ob_{ij}(\Rb^\alpha, \Rb^\beta) = \frac{1}{8 \pi \eta d}\left(
  \delta_{ij} + \frac {d_i d_j}{d^2}\right),
  \label{oseen_free}
\end{equation}
with ${\bf d} \equiv \Rb^\alpha- \Rb^\beta$ and $d \equiv |{\bf d}|$.
Here, the viscosity of the fluid $\eta$ is assumed to be equal for the
fluids inside and outside the vesicle \cite{fnt_viscosity}.

After dealing with the energetics of the membrane in the force balance
we have to treat the dynamics of the lipid membrane correctly.
Two-dimensional incompressibility of the membrane leads to the
condition that the area around every vertex remains constant under the
dynamics, i.e.~$0 = \partial_t A^\alpha = \sum_\beta (\partial
A^\alpha / \partial \Rb^\beta) \vb(\Rb^\beta)$.  Using
Eq.~(\ref{veloc}) for the velocities and Eq.~(\ref{force}) for the
vertex forces, this condition yields an inhomogeneous system of linear
equations for the local tensions $\sigma^\alpha$. After the tensions
are thus determined, moving the vertices by $\Delta
t\,\vb(\rb^\alpha)$ with a short timestep $\Delta t$ respects the
Stokes equations and all boundary conditions and thus gives a correct
integration scheme for the hydrodynamic problem which automatically
conserves the volume of the enclosed fluid.  Fluidity of the membrane
is ensured by bond flips attempted in regular intervals
\cite{s:krol92}.

This procedure turns out to be stable in the sense that (i) an
initially smooth surface remains smooth during time evolution, and
(ii) the distribution of triangle sizes and angles remains
approximately constant in the stationary state. As a test, we relax
surfaces at $\vb^0 = 0$ and arrive at the known axisymmetric
equilibrium shapes \cite{seif95b} with an error of $0.2$ percent for
the energy and $1.5$ percent for the (uniform) tension using a
discretization with $N=337$ vertices.  Area and volume remain constant
with an error below $0.1$ percent for the longest runs.


The numerical results for shape, orientation and membrane velocity in
the stationary state may be understood using a simplified model.  The
numerical data show an approximate mirror symmetry of the induced
tension $\sigma$ with respect to the plane defined by the axis denoted
by $a$ in Fig.~\ref{fig:shearscheme} (a) and the $y$-axis.  The axis
perpendicular to this symmetry plane is denoted by $c$.  We now assume
that this symmetry holds exactly.  The velocity field can be written
as a superposition of the undisturbed shear flow $\vb^0$ and the flow
$\vb^i$ induced by all forces on the membrane.  The latter flow must
have the mirror symmetry of the inducing forces.  In a stationary
state, the sum of these two contributions has to be tangential to the
membrane surface. We evaluate this condition for the contour of the
vesicle in the $x$-$z$-plane at $y=0$.  With $s$ denoting the
arclength of this contour, the contour is described by a vector
$(a(s), 0, c(s))$ with the local tangential velocity $\beta(s)$.  We
obtain four equations
\begin{eqnarray}
  \label{velsumI}
  \!\pm\beta(s)\frac{d a(s)}{d s} & = &
  v^i_a + \gamdot \cos\theta\left ( a(s)\sin\theta 
   \pm c(s)\cos\theta\right), \\
  \beta(s)\frac{d c(s)}{d s} & = & \pm
  v^i_c - \gamdot \sin\theta\left ( a(s)\sin\theta
   \pm c(s)\cos\theta\right),\label{velsumII}
\end{eqnarray}
where the different signs apply to the upper and to the reflected point
on the lower part of the vesicle, respectively.

We can now eliminate the unknown quantity $\vb^i$ from (\ref{velsumI})
- (\ref{velsumII}) and obtain differential equations for $d\,a(s)/
ds$ and $d\,c(s)/ds$.  These equations describe an ellipse
\begin{equation}
  \label{shear_ellips}
  \frac{a^2(s)}{r^2\;\cos^2\theta} + 
  \frac{c^2(s)}{r^2\;\sin^2\theta} = 1.
\end{equation}

If one now assumes axisymmetry of the entire \emph{shape}, the length
scale $r$ and the angle $\theta$ are uniquely determined by the
constraints on area and volume of the vesicle. The shear rate
$\gamdot$ scales the velocity $\beta(s)$ and does not influence the
shape or orientation. We obtain the differential equation
\cite{fnt_beta_sign}
\begin{equation}
  \label{angle_dgl_beta}
  \frac{d \beta(s)}{d s} = \pm \frac{\gamdot\,a(s)\,c(s)\,\sin^2\theta\,
    \cos^2\theta\,
    (\cos^2\theta-\sin^2\theta)}
  {a^2(s)\,\sin^4\theta + c^2(s)\,\cos^4\theta}
\end{equation}
for the velocity along the contour by forming the derivative of the
geometrical condition $(d\,a(s)/ds)^2 + (d\,c(s)/ds)^2 = 1$ with
respect to $s$.  This equation can be solved numerically, giving the
revolution frequency $\omega \equiv 2 \pi / \oint \beta(s)^{-1}\,ds$.

In the spherical limit $v \approx 1$, Eq.~(\ref{shear_ellips}) gives
$\theta \approx \pi / 4$ and $r \approx R_0$.
Eq.~(\ref{angle_dgl_beta}) simplifies to $\beta(s) = \gamdot R_0 / 2 =
\mathit{const}$, which is equivalent to $\omega = \gamdot / 2 = \chi /
2 \tau$.  Thus, all limit values for a spherical vesicle are identical
to the results for a rigid sphere or a fluid drop with infinite
surface tension in shear flow \cite{vdve89}.


As Figs.~\ref{fig:angle} and \ref{fig:tt} show, our simplified model
yields good quantitative agreement with the results of the
hydrodynamic integration. The results for different shear rate $\chi$
collapse as expected.  The spherical limit $v \approx 1$ can not be
reached numerically, as the tension in the membrane diverges when
exterior forces are applied to a sphere.  For reduced volumes $v \lsim
0.8$, the symmetry assumptions in our simplified model are less
justified.

\begin{figure}[tb]
  \begin{center}
    \leavevmode
    \psfig{file=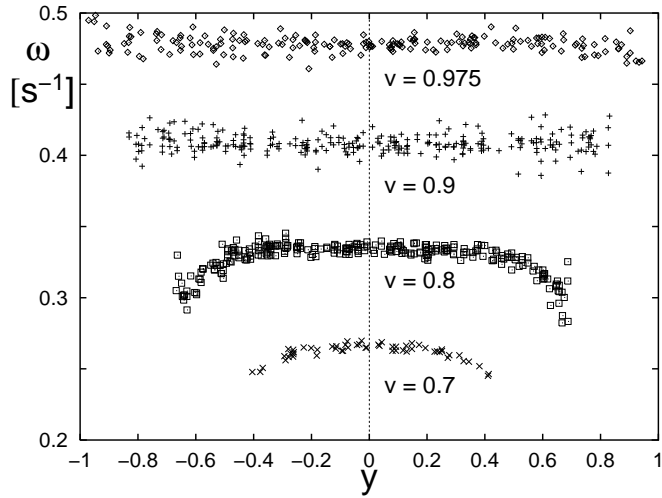,width=9.5cm}
  \end{center}
  \caption{
    Revolution frequency $\omega$ (as a function of lateral coordinate
    $y$ for reduced volumes $v=0.7, 0.8, 0.9, 0.975$ for $\chi = 10$,
    $\tau = 10 {\rm s}$. The $y$-independence of $\omega$ for large
    $v$ implies that adjacent membrane segments are again close
    together after a full revolution.}
  \label{fig:shear_v}
\end{figure}

The results of the numerical integration contain information about the
distribution of velocity and tension within the membrane not available
from the simplified model.  The revolution frequency
$\overline{\omega}$ only measures the mean velocity. The local
variations of membrane velocity show two remarkable features: (i) The
velocity varies along the contour in a way that velocities are smaller
towards the poles of the vesicle.  This behavior can be understood
qualitatively by formulating (\ref{velsumI}) and (\ref{velsumII}) for
a contour rotated out of the $x$-$z$ plane. (ii)
Fig.~\ref{fig:shear_v} shows the lateral \mbox{($y$-)}variation in the
revolution frequencies $\omega$ evaluated for individual vertices.
Even though we have neither included an explicit shear rigidity nor
shear viscosity, we find numerically effective shear in the membrane
only for large deviations from an elliptical shape at small $v$.

In the stationary state, the membrane tension $\sigma^\alpha$ is
constant only for $\vb^0 = 0$. In a shear flow, we find that the
largest values of the (negative) tension are in the middle of the
vesicle, i.e.~the vesicle responds to being pulled apart by the shear
flow. For large shear rates $\chi \gg 1$, bending rigidity is irrelevant,
and the mean tension $N^{-1} \sum_\alpha \sigma^\alpha$ increases
linearly with $\chi$, thus dominating the elastic energy of the
membrane.

In conclusion, we have developed a stable numerical scheme for
calculating the time evolution of vesicles with arbitrary shape
suspended in arbitrary flow fields. In the case of simple shear flow,
we find an inclination angle $\theta$ independent of shear rate and a
revolution frequency $\omega$ linearly proportional to shear rate.
Our approach breaks down for shear rates smaller than $\chi \approx
k_{\rm B}T / \kappa \simeq 0.04$, where typical velocities of
rotational diffusion $k_{\rm B}T / \eta R_0^2$ are comparable to the
velocity of the shear flow. In this regime, one should incorporate
thermal fluctuations.

We thank T.M.~Fischer for helpful discussions.  M.K.~and W.W.~thank
K.~Kehr from IFF, Forschungszentrum J\"ulich for hospitality.
U.S.~benefitted from interaction with S.~Langer and M.~Wortis on a
related problem.

\pagebreak


\begin{thebibliography}{10}

\bibitem{seif95b}
U. {Seifert} and R. {Lipowsky},  in {\em Structure and Dynamics of Membranes},
  edited by R. {Lipowsky} and E. {Sackmann} (Elsevier Science, Amsterdam,
  1994).

\bibitem{fisc78}
T.~M. {Fischer}, M. {St{\"o}hr--Liesen}, and H. {Schmid--Sch{\"o}nbein},
  Science {\bf 24},  894  (1978).

\bibitem{gron80}
W. {Groner}, N. {Mohandas}, and M. Bessis, Clin. Chem. {\bf 26},  1435  (1980).

\bibitem{pete92b}
M.~A. {Peterson}, Phys. Rev. A {\bf 45},  4116  (1992).

\bibitem{kell82}
S.~R. {Keller} and R. {Skalak}, J. Fluid Mech. {\bf 120},  27  (1982).

\bibitem{bart91}
D. {Barth\`es-Biesel}, Physica A {\bf 172},  103  (1991).

\bibitem{uijt93}
W.~S.~J. {Uijttewaal}, E.~J. {Nijhof}, and R.~M. {Heethaar}, Phys. Fluids A
  {\bf 5},  819  (1993).

\bibitem{pozr95}
C. {Pozrikidis}, J. Fluid Mech {\bf 297},  123  (1995).

\bibitem{broc75a}
F. {Brochard} and J.~F. {Lennon}, J. Physique {\bf 36},  1035  (1975).

\bibitem{miln87}
S.~T. {Milner} and S.~A. {Safran}, Phys. Rev. A {\bf 36},  4371  (1987).

\bibitem{nels95}
P. {Nelson}, T. {Powers}, and U. {Seifert}, Phys. Rev. Lett. {\bf 74},  3384
  (1995).

\bibitem{folt94}
G. {Foltin}, Phys. Rev. E {\bf 49},  5243  (1994).

\bibitem{cai95}
W. {Cai} and T.~C. {Lubensky}, Phys. Rev. E {\bf 52},  4251  (1995).

\bibitem{fnt_c0}
We employ the simplest possible curvature model for the membrane bending
  energy, neglecting spontaneous curvature of the membrane as well as effects
  of bilayer elasticity \cite{seif95b}. These would introduce additional
  length-scales.

\bibitem{fnt_numtt}
Numerically, we determine the mean tank-treading time $t_t = 2 \pi / \omega$ as
  twice the time between two successive sign changes of the coordinate $z$ for
  a single vertex and average over all vertices in order to obtain
  $\overline{\omega}$.
  
\bibitem{krau96} The smallest $\chi$ for which we could obtain
  numerically stable data is $\chi = 0.1$, cf.~M. {Kraus}, Ph.D.
  thesis, {Universit\"at Potsdam}, 1996.

\bibitem{rall84}
J.~M. {Rallison}, Ann. Rev. Fluid Mech. {\bf 16},  45  (1984).

\bibitem{krau95z}
M. {Kraus}, U. {Seifert}, and R. {Lipowsky}, Europhys. Lett. {\bf 32},  431
  (1995).

\bibitem{vdve89}
T.~G.~M. {van de Ven}, {\em Colloidal Hydrodynamics} (Academic Press, London,
  1989).

\bibitem{fnt_curvdiscr}
Squared mean curvature is discretized by calculating the mean curvature on the
  edges of the triangulation and subsequently summing up and squaring on
  adjacent vertices.

\bibitem{canh70}
P.~B. {Canham}, J. Theoret. Biol. {\bf 26},  61  (1970).

\bibitem{helf73}
W. {Helfrich}, Z. Naturforsch. {\bf 28c},  693  (1973).

\bibitem{pozr92}
C. Pozrikidis, {\em Boundary Integral and Singularity Methods for Linearized
  Viscous Flow} (Cambridge University Press, Cambridge, 1992).

\bibitem{doi86}
M. {Doi} and S.~F. {Edwards}, {\em The Theory of Polymer Dynamics} (Clarendon
  Press, Oxford, 1986).

\bibitem{fnt_oseendiv}
The Oseen tensor diverges for $d \to 0$. The self-interaction term $\Ob^{{\rm
  f}} (\Rb^\alpha,\Rb^\alpha)$ can be regularized by integrating over the
  adjacent triangles of $\Rb^\alpha$, cf.~\cite{uijt93}. As this integration
  procedure is employed only for this special case and not for the interaction
  between arbitrary vertices, we have to introduce a weighting factor
  $c^\alpha$ for the self-interaction which is chosen as to ensure constancy of
  the vesicle volume. Usually, we find \mbox{$\max_\alpha | 1 - c^\alpha | <
  0.05$}.

\bibitem{fnt_viscosity}
Different viscosities inside and outside the vesicle membrane lead to an
  integral equation in $\vb$ for every time-step, cf.~Ref.~\cite{pozr92}.

\bibitem{s:krol92}
D.~M. {Kroll} and G. {Gompper}, Science {\bf 255},  968  (1992), and references
  therein.

\bibitem{fnt_beta_sign}
Positive sign applies to $a c<0$, and negative sign to $a c >0$.

\end{thebibliography}
\end{document}